\documentclass[10pt,prl,aps,twocolumn,superscriptaddress,preprintnumbers]{revtex4-1}

\usepackage[utf8]{inputenc}
\usepackage{epigraph}
\usepackage{amsmath}
\usepackage{wrapfig}
\usepackage{epsfig}
\usepackage{amssymb}
\usepackage{graphicx}
\usepackage{slashed}
\usepackage{xcolor}
\usepackage{comment}
\usepackage{natbib}
\usepackage{enumitem}
\usepackage{tikz-cd,tikz}
\usepackage{float}
\usepackage{array,adjustbox,booktabs}
\usepackage{subcaption}
\usepackage{amsfonts}
\usepackage{physics}
\usepackage{hyperref}

\usepackage{xcolor}
\usepackage{colortbl}
\definecolor{dark-red}{rgb}{0.80,0.12,0.12} % links
\definecolor{dark-blue}{rgb}{0,0.15,0.85} % citelink

\hypersetup{
  linktocpage,
  colorlinks  = true, %Colours links instead of ugly boxes
  urlcolor    = magenta, %Colour for external hyperlinks
  linkcolor   = dark-blue, %Colour of internal links
  citecolor   = purple %Colour of citations
}

\usepackage[a4paper]{geometry}
\DeclareMathAlphabet\mathbfcal{OMS}{cmsy}{b}{n}

\usetikzlibrary{decorations.markings}
\tikzset{middlearrow/.style={
		decoration={markings,
			mark= at position 0.5 with {\arrow{#1}} ,
		},
		postaction={decorate}
	}
}

\newcommand{\arxiv}[1]{\href{https://arxiv.org/abs/#1}{arXiv:#1}}

\def\be{\begin{equation}}
\def\ee{\end{equation}}
\def\eqn#1{eq.~\eqref{#1}}
\def\eqns#1#2{eqs.~\eqref{#1} and \eqref{#2}}
\def\rcite#1{ref.~\cite{#1}}
\def\rcites#1{refs.~\cite{#1}}

\def\redA{\hat{\mathcal{A}}}

\def\intsum{\mathop{\sum \kern-1.2em \int}\limits_{\check{s}}d \hat{s}}

\def\intsumuno{\mathop{\sum \kern-1.2em \int}\limits_{\sum_j\!\check{s}_{ij}=\,p_i-1}\kern-1.5em{d} \hat{s}}
\def\intsumdue{\mathop{\sum \kern-1.2em \int}\limits_{\sum_j\!\check{s}_{ij}=\,p_i-2}\kern-1.5em{d} \hat{s}}
\def\intsumtre{\mathop{\sum \kern-1.2em \int}\limits_{\sum_j\!\check{s}_{ij}=\,p_i-3}\kern-1.5em{d} \hat{s}}
\def\omegatwo{\Omega_2}
\def\tree{\begin{tikzpicture}[scale=0.05]

  \draw[fill=brown!70!black] (0,0) rectangle (0.6,3);

  \begin{scope}[shift={(0.3,3)}] % Center canopy over trunk
    \draw[fill=green!60!black] (0,0.5) circle (1.4);  % Center top
    \draw[fill=green!60!black] (-1,0) circle (1.1);   % Left
    \draw[fill=green!60!black] (1,0) circle (1.1);    % Right
    \draw[fill=green!60!black] (0,-0.5) circle (1);   % Bottom
  \end{scope}

\end{tikzpicture} }

\def\oneloop{\begin{tikzpicture}[scale=0.06]
\def\shift{0.1};
\def\shiftt{0.2};
\draw[thick,dashed] (-1.2,-0.5)--(0.33,0.6);
\draw[thick,dashed] (-1.2,2.5)--(0.33,1.6);
\draw[thick,dashed] (4.5-3,0.5)--(6.3-3,-0.6);
\draw[thick,dashed] (4.5-3,1.6)--(6.3-3,2.5);
\fill[white] (1,1) circle (1);
\draw[thick] (1,1) circle (1);
\end{tikzpicture}
}

\newcommand{\nocontentsline}[3]{}
\let\origcontentsline\addcontentsline
\newcommand\stoptoc{\let\addcontentsline\nocontentsline}
\newcommand\resumetoc{\let\addcontentsline\origcontentsline}

\begin{document}

\preprint{
UUITP-29/25,
QMUL-PH-25-31
}

\title{Surprising one-loop finiteness of 6D half-maximal supergravities}
\author{Yu-tin Huang}
\email{yutin@phys.ntu.edu.tw}
\affiliation{Department of Physics and Center for Theoretical Physics, National Taiwan University, Taipei 10617, Taiwan}
\affiliation{Physics Division, National Center for Theoretical Sciences, Taipei 10617, Taiwan}
\affiliation{Max Planck{-}IAS{-}NTU Center for Particle Physics, Cosmology and Geometry, Taipei 10617, Taiwan}
\author{Henrik Johansson}
\email{henrik.johansson@physics.uu.se}
\affiliation{Department of Physics and Astronomy, Uppsala University, Box 516, 75120 Uppsala,
Sweden}
\author{Michele Santagata}
\email{michele.santagata@physics.uu.se}
\affiliation{Department of Physics and Center for Theoretical Physics, National Taiwan University, Taipei 10617, Taiwan}
\affiliation{Department of Physics and Astronomy, Uppsala University, Box 516, 75120 Uppsala,
Sweden}
\author{Congkao Wen}
\email{c.wen@qmul.ac.uk}
\affiliation{Centre for Theoretical Physics, Department of Physics and Astronomy,
Queen Mary University of London, London, E1 4NS, UK}
\begin{abstract}
\noindent 
In four dimensions, it has long been established that gravity coupled to matter exhibits ultraviolet divergences at one loop, irrespective of supersymmetry. Notably, the four-matter one-loop amplitudes of half-maximal supergravity coupled to Maxwell multiplets were shown in the 1970s to be divergent. Surprisingly, we demonstrate in this work that half-maximal theories can nevertheless become one-loop finite when uplifted to higher dimensions, contrary to naive expectations. Specifically, we study the ultraviolet properties of the four-matter and two-matter two-graviton amplitudes in six-dimensional $\mathcal{N}=(2,0)$ and $\mathcal{N}=(1,1)$  supergravities, coupled to $n_T$ tensor and $n_V$ vector multiplets, respectively. We find that the one-loop amplitudes are finite for $n_T=21$ and $n_V=20$. This finiteness is unexpected, as symmetry-preserving counterterms do exist. Interestingly, both  values exactly correspond to low-energy limits of type II string theories compactified on K3, 
which hints at possible origins to the surprising cancellations. 
\end{abstract}

\maketitle

\stoptoc
\section{Introduction}
Determining the fate of supergravity ultraviolet (UV) divergences is central to our understanding of the limits where the model in question ceases to be an effective field theory. Divergences indicate the appearance of new degrees of freedom, such as those of string theory. Conversely, absence of divergences due to unexpected cancellations, might hint at hidden symmetries or dualities. 

The UV behavior of four-dimensional pure supergravity theories remains unsettled (see~\rcites{Bern:2013uka, Bern:2018jmv} for the current status). However, when coupled to matter, these theories have been long known to diverge, regardless of the amount of supersymmetry or matter multiplets~\cite{Deser:1974cz, Fischler:1979yk,tHooft:1974toh}.  For $\mathcal{N}=4$ supergravity coupled to $n_V$ Maxwell multiplets the divergence is proportional to $(1+n_V/2) F^4$~\footnote{The counterterm is $(T_{\mu\nu})^2=(F_{\mu\rho}F_{\nu}\,^\rho-\frac{1}{4}(F_{\alpha\beta})^2\eta_{\mu\nu})^2$.}, and thus no amount of matter can render it finite. 

In general, if a theory is divergent in a given dimension, it is usually expected to be divergent in higher dimensions. Thus, one would naively expect half-maximal supergravity theories with arbitrary matter multiplets be divergent in six dimensions as well. This expectation is also supported by the fact that symmetry-preserving counterterms with one-loop power counting exist, as we will discuss later.

In this Letter, we study the $\mathcal{N}=(2,0)$ chiral supergravity theory coupled to $n_T$ abelian self-dual tensor multiplets as well as $\mathcal{N}=(1,1)$ non-chiral supergravity theory coupled to $n_V$ vector multiplets. In particular, we bootstrap the one-loop matter amplitudes from tree-level ones by combining constraints from two-particle unitary cut and crossing symmetry (see~\rcite{Huang:2024ihm} for a similar strategy in the context of string amplitudes). 
To our surprise, we find that the coefficient in front of the scale-dependent logarithmic terms, which reflects the presence of UV divergence, is proportional to $(n_T{-}21)$ for the $\mathcal{N}=(2,0)$ system and $(n_V{-}20)$ for $\mathcal{N}=(1,1)$, making the two theories finite at one loop for critical values $n_T=21$ and $n_V=20$, respectively. 
As an independent check, we analyze the UV properties of the one-loop four-matter amplitudes using the BCJ double-copy construction~\cite{Bern:2010ue}. This method has the additional advantage of extending naturally to amplitudes involving external gravitons, and we employ it to study the UV behavior of the two-matter–two-graviton amplitude. Using dimensional regularization, we find that the resulting $1/\epsilon$ divergences display the same remarkable cancellation.

Notably, the condition $n_T=21$ coincides with the well-known requirement for cancellation of the six-dimensional gravitational anomaly~\cite{Erler:1993zy}, although we are not aware of any direct argument linking anomaly cancellation to UV finiteness.
Even more intriguingly, these two finite setups are realized precisely in type II string theories compactified on K3~\cite{Aspinwall:1996mn}. The low-energy limit of type IIB yields $\mathcal{N}=(2,0)$ supergravity with 21 tensor multiplets, while type IIA  (or alternatively heterotic string compactified on $T^{4}$) yields $\mathcal{N}=(1,1)$ supergravity with 20 vector multiplets. Upon further compactification on $S^1$, the theories are related by string dualities. It is therefore tempting to conjecture that the coincidence of their one-loop four-particle amplitudes is a low-energy reflection of the string dualities.

\section{Tree-level and on-shell setup}
We consider amplitudes of 6D supergravity with (2,0) and (1,1) supersymmetry, coupled to $n_T$ tensor multiplets and $n_V$ vector multiplets, respectively. 
Since these theories differ by a chirality flip in half of the supercharges, altering the discussion only slightly, we focus on the chiral (2,0) case and revisit the non-chiral (1,1) at the end.

The on-shell content of 6D (2,0) supergravity is organized into graviton and (self-dual) tensor  superfields,
\begin{align} \label{Multiplets}
     H_{\hat{a}\hat{b}}(\eta) &= B_{\hat{a}\hat{b}}+ \ldots + \eta_{a}^I \eta_{I,b} G_{\hat{a}\hat{b}}^{ab} + \ldots + (\eta)^4 \bar{B}_{\hat{a}\hat{b}}\, , \nonumber\\
    T(\eta) &= \phi + \ldots +\eta_{a}^I \eta_{I,b} B^{ab} + \ldots +  (\eta)^4 \bar{\phi} \,,
\end{align}
where the ellipses represent  suppressed fermionic and other bosonic states. 
The $\eta$'s are Grassmann variables and the unhatted (hatted) lowercase indices belong to the left(right)-handed $SU(2)_L\times SU(2)_R$ little group.  The tensor multiplet is a little-group  singlet, while the graviton multiplet $H_{\hat{a}\hat{b}}$  transforms under $SU(2)_{R}$ as a triplet. The index $I=1,2$ manifests a $SU(2)$ subgroup of the R-symmetry $USp(4)$.

A four-point (2,0) amplitude has the loop expansion
\begin{equation}\label{ampl_expans}
 \mathcal{A} = \ell_P^4\, \delta^{(8)}(Q) \big(  \redA^{\tree}  +   \ell_P^4 \redA^{\oneloop} +\cdots \big) \, , 
\end{equation}
where a \textit{reduced amplitude} can be defined by stripping off a supermomentum delta function $\delta^{(8)}(Q)$. 
The reduced tree-level and one-loop amplitudes are denoted by $\redA^{\tree}$ and $\redA^{\oneloop}$, respectively. Using the 6D spinor-helicity formalism~\cite{Cheung:2009dc}, the massless momentum of the $i$-th particle can be expressed as
\begin{equation}
    p^{AB}_i= \lambda_{i,a}^A  \lambda_{i,b}^B  \epsilon^{ab} = \frac{1}{2} \epsilon^{ABCD} \tilde{\lambda}_{i, C \hat{a}}\tilde{\lambda}_{i, D \hat{b}} \, \epsilon^{\hat{a} \hat{b}} \, ,
\end{equation}
where $\lambda$'s ($\tilde{\lambda}$'s) are on-shell spinor variables, and $A,\ldots, D$ are $Spin(5,1)\cong SU(2,2)$ indices of the Lorentz group. The supermomentum delta function can then be given in spinor-helicity variables, with positive energy states,  as $\delta^{(8)}(Q) = \delta^8 (\sum_{i=1}^2 \lambda_{i,a}^A \eta_i^{I, a}-\sum_{i=3}^4 \lambda_{i,a}^A \eta_i^{I, a})$.
Focusing on four-point matter superamplitudes, the tree-level pure-tensor $(TT)$ and mixed tensor-graviton $(TH)$ reduced amplitudes are \cite{Lin:2015dsa, Heydeman:2018dje}:
\begin{align}
\label{4tensor_tree}
& \redA_{TT}^{\tree; f_1 f_2 f_3 f_4}  =  \frac{\delta^{f_1 f_2}\delta^{f_3 f_4}}{s}{+}\frac{\delta^{f_1 f_4}\delta^{f_2 f_3}}{t}{+}\frac{\delta^{f_1 f_3}\delta^{f_2 f_4}}{u} \,, \\
\label{tensgrav_tree}
& \redA_{TH}^{\tree; f_1 f_2}= \delta^{f_1 f_2} \frac{[ 3_{(\hat{a}_3}|p_1|4_{(\hat{a}_4}]\, [ 3_{\hat{b}_3)}|p_1|4_{\hat{b}_4)} ]}{stu} \,.
\end{align}
The Mandelstam invariants $s=2 p_1 \cdot p_2$, $t=-2 p_2 \cdot p_3$, $u=-2 p_1\cdot p_3$ satisfy $s+t+u=0$, the anti-chiral  bracket is $[ i_{\hat{a}}|p_k|j_{\hat{b}}] = \tilde{\lambda}_{i,A\hat{a}} p_k^{AB}\tilde{\lambda}_{j,B\hat{b}}$,
and the $f_i$'s are flavor indices of $SO(n_T)$. For simplicity, we suppress all indices but flavor on the $\redA$'s. 
With the tree-level amplitudes at hand, we will now bootstrap the one-loop amplitude $\redA_{TT}^{\oneloop}$  
and study its UV properties.

\section{One-loop discontinuities}
Let us start by constructing the $s$-channel discontinuity by sewing together tree-level amplitudes and integrating over the on-shell phase space \footnote{We adopt the conventions of \rcite{Huang:2024ihm}, with the $s$-channel discontinuity defined as the difference above and below the branch cut, normalized as $\text{Disc}_{s} \log(-s)=1$. For a recent review on unitarity cuts, see \rcite{Britto:2024mna}.}.
Suppressing for simplicity the external indexes, we have
\begin{align}\label{discA}
& \text{Disc}_s[ \mathcal{A}_{TT}^{\oneloop}] =   \\
& = - \frac{s}{4\pi} \!\int d\omegatwo \!\!\!\!\!\sum_{X \in \{ T,H \}} \!\!\!\! \mathcal{A}_{TX}^{\tree}({1,2,\ell_1,\ell_2})   \mathcal{A}_{XT}^{\tree}({\ell_1,\ell_2,3,4})\, , \notag
\end{align}
where the sum runs over all intermediate states -- the tensor and graviton multiplets --
and $d\omegatwo$ is the 6D Lorentz-invariant two-particle phase space measure \cite{Soldate:1986mk},
\begin{equation}
  d\omegatwo = -\frac{1}{2(4\pi)^3} \sin^2{\theta_{\text{in}}} \sin^2{\phi_{\text{fn}}}   \,\, d\phi_{\text{fn}}\, d\cos \theta_{\text{in}} \,.
\end{equation}
In the center-of-mass frame, $\theta_{\text{in}}$, $\theta_{\text{fn}}$, are the angles between the incoming momentum $p_1$ and the loop momentum $\ell_1$, and between the outgoing momentum $p_3$ and $\ell_1$, respectively.
The angle $\phi_{\text{fn}}$ denotes the azimuthal separation between the initial and final scattering planes, while $\theta$ is the standard scattering angle between the incoming and outgoing particles $p_1$ and $p_3$, with $\cos\theta = 1+ 2t/s$. 
These angles are related by
 $\cos \theta_{\text{fn}} = \cos \theta \cos \theta_{\text{in}}+ \sin \theta \sin \theta_{\text{in}} \cos \phi_{\text{fn}}$.
%The configuration of the scattering process and the definitions of the various angles are illustrated in Fig.~\ref{fig_scattangle}. 
The invariants associated with the loop momenta, $s_{i\ell_j} =2 p_i \cdot \ell_j$, can be given in terms of the scattering angles
\begin{align}
   & s_{1\ell_1}=  \frac{s}{2}\big( 1-\cos \theta_{\text{in}}  \big)\,, \quad   s_{1\ell_2}=  \frac{s}{2}\big( 1+\cos \theta_{\text{in}}  \big)\,, \notag \\
   & s_{3\ell_1}=  \frac{s}{2}\big( 1-\cos \theta_{\text{fn}}  \big)\,, \quad  s_{3\ell_2}=  \frac{s}{2}\big( 1+\cos \theta_{\text{fn}}  \big) \,.
\end{align}

Intermediate tensor or graviton states propagating across the cut give the two contributions
\begin{align}
\label{disconti_decomp2}
\text{Disc}_{s}[\mathcal{A}_{TT}^{\oneloop; f_1 f_2 f_3 f_4}]& =\delta^{(8)}(Q)\big(\mathcal{T}^{f_1 f_2 f_3 f_4}+\mathcal{H}^{f_1 f_2 f_3 f_4} \big)\,,\nonumber \\
     \mathcal{T}^{f_1 f_2 f_3 f_4} &=-\frac{s^3}{4\pi} \int d\omegatwo  \redA_{TT}^{\tree; f_1 f_2 f_5 f_6}   \redA_{TT}^{\tree; f_5 f_6 f_3 f_4}  \, , \notag \\ 
         \mathcal{H}^{f_1 f_2 f_3 f_4} &=- \frac{s^3}{4\pi} \int d\omegatwo   \redA_{TH}^{\tree; f_1 f_2}   \redA_{HT}^{\tree; f_3 f_4}   \, .
\end{align}
We used the fact that the supermomentum self-replicates
\begin{equation}\label{sewing_rel}
\sum_{\ell_{1},\ell_{2}\text{ states}} \delta^{(8)}(Q_L) \delta^{(8)}(Q_R) = s^2 \delta^{(8)}(Q)\, ,
\end{equation}
where $Q=Q_L+Q_R$, and the latter are the supermomenta of the left and right amplitudes in \eqn{discA}. Thus, the reduced amplitudes are glued on the cut simply by including an additional factor of $s^2$. We now separately analyze the tensor and graviton contributions:
\vspace{3pt}

{\bf{Tensor cut.}} 
The contribution from the tensor cut is obtained by multiplying two reduced pure-tensor amplitudes \eqref{4tensor_tree}, giving the tensor discontinuity
\begin{align} \label{eqn11}
  &   \mathcal{T}^{f_1 f_2 f_3 f_4}\! =\!   \big(\delta^{f_1f_2}\delta^{f_3 f_4} I_s  {+} \delta^{f_1f_4}\delta^{f_2 f_3} I_t{+}\delta^{f_1f_3}\delta^{f_2 f_4} I_u \big), \nonumber \\
&I_s = \frac{s^2}{4\pi} \int d\omegatwo \Big(\frac{1}{s_{3\ell_1}}{+}\frac{1}{s_{3\ell_2}}{+}\frac{1}{s_{1\ell_1}}{+}\frac{1}{s_{1\ell_2}}{-}\frac{n_T}{s} \Big)\,, \\
&I_t = -\frac{s^3}{4\pi} \int d\omegatwo\Big( \frac{1}{s_{1\ell_2}s_{3\ell_1}}{+}\frac{1}{s_{1\ell_1}s_{3\ell_2}}\Big)\,,~~~~I_u =I_t\Big|_{\theta \rightarrow \theta+\pi}\,. \nonumber
\end{align} 
The integrals can be evaluated explicitly, yielding
\begin{align}
& I_s =- \frac{s}{(4\pi)^3}   \frac{n_T-12 }{12}  \,,~~~ \\
& I_t  =  \frac{1}{(4\pi)^3} \frac{s^2}{u}  \log \big({-}\frac{s}{t}\big)\,,~~~~ I_u  =  \frac{1}{(4\pi)^3 }  \frac{s^2}{t} \log \big({-}\frac{s}{u}\big)\,. \notag 
\end{align}
\vspace{3pt}

 {\bf{Graviton cut.}}
Let us now consider the case when the intermediate states are gravitons. This can be readily obtained by inserting two tensor-graviton amplitudes \eqref{tensgrav_tree} on either side of the cut. After some straightforward manipulations, one arrives at
\begin{align}\label{disc_grav}
 & \mathcal{H}^{f_1 f_2 f_3 f_4}=-\delta^{f_1f_2}\delta^{f_3 f_4} \frac{s}{8\pi} \! \int \! \frac{{\cal I}_{H}^{(2,0)}}{ s_{1\ell_1}s_{1\ell_2}s_{3\ell_1}s_{3\ell_2}}  d\omegatwo \, ,
\end{align}
where the numerator is given in terms of 6D Dirac traces
\begin{align}
\label{disco_grav}
{\cal I}_{H}^{(2,0)} =& \Tr^2 ( p_1 \ell_2  p_3 \ell_1){+} s_{1\ell_2}s_{3\ell_1} \! \Tr ( p_3  \ell_1  p_1 \ell_2) \notag \\
&{-}s_{1\ell_1}s_{3\ell_2} \Tr (p_1 \ell_2  \ell_1 p_3) {+}s \, u \Tr (\ell_1 p_1 \ell_2 p_3) \, .
\end{align}
Upon integration, one obtains
\begin{equation} \label{GravitonFunction}
 \mathcal{H}^{f_1 f_2 f_3 f_4}=    \frac{ \delta^{f_1f_2} \delta^{f_3 f_4}}{(4\pi)^3}  \Big[ \frac{t^2}{ u} \!\log \big({-}\frac{s}{t}\big)   +\frac{u^2}{ t} \!\log \big( {-}\frac{s}{u}\big)   + \frac{3}{4} s \Big].
\end{equation}
Combining the tensor and graviton contributions gives
\begin{equation}
    \text{Disc}_{s}[\redA_{TT}^{\oneloop}] \!=\!  \delta^{f_1f_2}\delta^{f_3 f_4} \mathcal{V}_s + \delta^{f_1f_4}\delta^{f_2 f_3}\mathcal{V}_t+\delta^{f_1f_3}\delta^{f_2 f_4}\mathcal{V}_u,  
\end{equation}
with
\begin{align}\label{CsCtCu}
    &   \mathcal{V}_s =   \frac{ 1}{(4\pi)^3}  \Big[ \frac{t^2}{ u} \log \big({-}\frac{s}{t}\big)   +\frac{u^2}{ t} \log \big({-}\frac{s}{u}\big)   - \frac{n_T-21}{12} s \Big]  \,,  \notag \\
        &   \mathcal{V}_t =   \frac{ 1}{(4\pi)^3}  \frac{s^2}{ u} \log \big({-}\frac{s}{t}\big)  \,, ~~~~ \mathcal{V}_u =\mathcal{V}_t\Big|_{t \leftrightarrow u}\,.
%    &   \mathcal{V}_u = -  \frac{ 1}{(4\pi)^3}  \frac{s^2}{ t} \log \big( {-}\frac{s}{u}\big) \,.
\end{align}

\section{One-loop amplitudes}
With the discontinuity in hand, we now proceed to reconstruct the one-loop amplitude. This is achieved by making an ansatz in terms of logarithmic functions and imposing a few physical constraints. All logarithmic contributions uniquely follow from the known discontinuities and crossing symmetry, while rational terms are fixed by requiring the absence of spurious singularities.

By construction, this approach is insensitive to polynomial contributions, which both are cut- and pole-free, and is therefore agnostic to any specific regularization scheme. We will later fix the remaining polynomial ambiguity by constructing the loop integrands explicitly via the double copy, and evaluating them using dimensional regularization.

\vspace{3pt}

{\bf 6D (2,0) supergravity.}
For simplicity, we display the final result in the unphysical region $s<0$; the expression valid in the physical region is then obtained via analytic continuation.
The four-matter one-loop amplitude in chiral supergravity reads
\begin{equation}\label{oneloopamp}
    \redA_{TT}^{\oneloop} =  \delta^{f_1f_2}\delta^{f_3 f_4}\!\mathcal{F}_{s}^{(2,0)}  + \delta^{f_1f_4}\!\delta^{f_2 f_3} \mathcal{F}_{t}^{(2,0)}+\delta^{f_1f_3}\delta^{f_2 f_4}\!\mathcal{F}_{u}^{(2,0)},
\end{equation}
where
\begin{align}\label{Btu}
\mathcal{F}_{s}^{(2,0)} \! = -  \frac{1}{(4\pi)^3} \Big[& t^2 \mathcal{B}_{s,t} +  u^2 \mathcal{B}_{s,u}  + \frac{n_T-21}{12} s \log\big({-}\frac{s}{\mu^2}\big) \Big], 
\end{align}
and $\mathcal{B}_{s,t}$ is the 6D scalar box diagram:
\begin{equation}
\mathcal{B}_{s,t} =\frac{1}{2}\frac{1}{s+t} \Big(\log^2 \frac{s}{t}+\pi^2 \Big)\,,
\end{equation}
while the other contributions are obtained by cyclic permutations of $s,t,u$. In \eqn{Btu}, we have introduced a renormalization scale $\mu$ to ensure that the argument of the logarithm is dimensionless. Upon analytic continuation, it is straightforward to check that \eqref{oneloopamp} reproduces the expected discontinuities, e.g. $\text{Disc}_{s}[\mathcal{F}_{s}^{(2,0)}] = \mathcal{V}_{s}$, and likewise for the other crossing channels. Moreover, the $\pi^2$ term is required to ensure the correct analytic structure after analytic continuation to the physical region.

Finally, for $n_T=21$, the coefficient of the $\log(-s/\mu^2)$ term vanishes, showing the amplitude is finite in this special case. 

\vspace{3pt}

{\bf 6D (1,1) supergravity.}
The analysis for 6D (1,1) supergravity proceeds analogously to the chiral case, so here we only highlight the key differences. The supermomentum delta function is replaced by $\delta^{4}(Q)\delta^{4}(\tilde{Q})$, the reduced four-matter amplitude \eqref{4tensor_tree} is unchanged, while \eqn{tensgrav_tree} is superseded by the vector-graviton amplitude
\begin{equation}
\label{tensgrav_tree11}
\redA_{VH}^{\tree; f_1 f_2}= \delta^{f_1 f_2} \frac{\big[ 3_{\hat{a}_3}|p_1| 4_{\hat{a}_4}\big]\langle 3_{a_3}|p_1| 4_{a_4} \rangle}{stu} \,,
\end{equation}
with a chiral bracket $\langle i_{a}|p_k|j_{b}\rangle = \lambda_{i,a}^A p_{k,AB}\lambda_{j,b}^B$. 
Thus, the one-loop analysis for the matter cut is unaltered, whereas the graviton cut for the (1,1) theory is
\begin{align}
%&  \tilde{I}_{H}^{(1,1)} = \Tr^2 ( p_1 \ell_2  p_3 \ell_1) \, ; \\[2pt]
&  -\delta^{f_1f_2}\delta^{f_3 f_4}\frac{s}{4\pi} \int \frac{\Tr^2 ( p_1 \ell_2  p_3 \ell_1) }{s_{1\ell_1}s_{1\ell_2}s_{3\ell_1}s_{3\ell_2}}  \,\, d\omegatwo \,  \\
& =    \frac{\delta^{f_1f_2}\delta^{f_3 f_4}}{(4\pi)^3} \Big[ \frac{t^2}{ u} \log \big({-}\frac{s}{t}\big)   +\frac{u^2}{ t} \log \big({-}\frac{s}{u}\big)   + \frac{2}{3} s \Big] \, . \notag  
\end{align}
The only minor difference compared to \eqn{GravitonFunction} is in the linear-in-$s$ term.
Combining vector and graviton contributions and bootstrapping the one-loop amplitude $\redA_{VV}^{\oneloop}$ as before, we recover the same structure as in \eqn{oneloopamp}, with $\mathcal{F}_{s}^{(1,1)}$ altered only for this linear-in-$s$ term
\begin{align}\label{Btu11}
\mathcal{F}_{s}^{(1,1)} \!= - \frac{1}{(4\pi)^3} \Big[& t^2 \mathcal{B}_{s,t} +  u^2 \mathcal{B}_{s,u}  + \frac{n_V-20}{12} s \log\big({-}\frac{s}{\mu^2}\big) \Big], 
\end{align}
with similar expressions for the other crossing channels. In this case, the $\log(-s/\mu^2)$ term vanishes when $n_V=20$.

\vspace{5pt}

\section{UV divergences from the double copy}
As shown above, the $\log(-s/\mu^2)$ coefficient cancels when the number of matter multiplets is tuned to special values. We now present an alternative proof of this cancellation by constructing the integrand via the double copy~\cite{Bern:2008qj,Bern:2010ue,Johansson:2014zca} and analyzing the UV divergence in dimensional regularization. 
Beyond providing a consistency check of the unitarity-based result for the four-matter amplitudes, this approach offers two additional advantages. First, it generalizes straightforwardly to the mixed matter-graviton amplitude, which -- as we will see -- exhibits the same cancellation. Second, it determines the local polynomial terms that cannot be extracted from the cut discontinuities.

An $L$-loop supersymmetric-QCD (SQCD) amplitude, in $D\le 6$ dimensions and with $n_{\Phi}$ fundamental massless hypermultiplets, can be written in the form~\cite{Johansson:2015oia,Johansson:2017bfl}
\begin{equation}
 \redA^{(L)}_{\rm SQCD}=  \sum_{i} \int \frac{d^{DL}\ell}{(2\pi)^{DL}}\frac{(n_\Phi)^{|i|}}{S_i}\frac{C_i N_i}{{\cal D}_i}\,,
\end{equation}
where the sum is over all cubic Feynman (super)graphs of SQCD. For each graph $i$, we associate a color factor $C_i$, a symmetry factor $S_i\in  \mathbb{N}$, an index $|i|\in  \mathbb{N}\cup \{0\}$ counting the number of hyper loops, a propagator-denominator factor ${\cal D}_i$, and finally a kinematic numerator $N_i=N_i(\ell)$.
The kinematic numerators are not given by SQCD Feynman rules, instead they are constrained to satisfy color-kinematics duality; enforcing a $\{i,j,k\}$-triplet kinematic numerator identity for each color identity,
\be
N_i-N_j=N_k ~~~ \Leftrightarrow ~~~ C_i-C_j=C_k\,.
\ee
The three-term color identities originate from either a Jacobi identity $f^{abe}f^{ecd}-f^{ace}f^{ebd}=f^{ade}f^{ecb}$  or a Lie algebra identity $[T^a,T^b]=if^{abc}T^c$.

\begin{figure}[t]
  \centering
  \begin{subfigure}[a]{0.079\textwidth}
    \centering
\includegraphics[width=\textwidth]{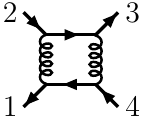}   
    \caption{}
  \end{subfigure}
  \hfill  
  \begin{subfigure}[a]{0.094\textwidth}
    \centering
\includegraphics[width=\textwidth]{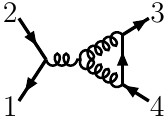}   
    \caption{}
  \end{subfigure}
  \hfill
  \begin{subfigure}[a]{0.094\textwidth}
    \centering
\includegraphics[width=\textwidth]{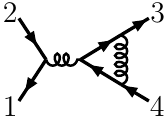}
    \caption{}
  \end{subfigure}
    \hfill
      \begin{subfigure}[a]{0.097\textwidth}
    \centering
\includegraphics[width=\textwidth]{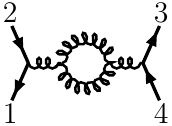}   
    \caption{}
  \end{subfigure}
\hfill  
\begin{subfigure}[a]{0.097\textwidth}
    \centering
\includegraphics[width=\textwidth]{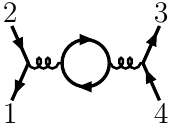}   
    \caption{}
  \end{subfigure}
  \caption{The BCJ diagram topologies for 6D SQCD with external hypermultipets. Arrows carry $USp(n_\Phi)$ flavor charge and numerators map $N_i\to-N_i$ under arrow flip.  }
  \label{fig:1loopBCJdiagrams}
\end{figure}

The double copy then allows us to construct $L$-loop gravitational amplitudes in half-maximal supergravities in $D\le 6$ dimensions, by replacing the color factors by corresponding numerators and also reinterpreting the $n_{\Phi}$ numbers~\cite{Johansson:2017bfl}. Depending on whether we use chiral numerators $N_i$, or anti-chiral $\tilde N_i$ (related by 6D parity), we get two options
\begin{equation}\label{DoubleCopyAnyL}
 \redA^{(L)}_{\rm SG}=  \sum_{i} \! \int \! \frac{d^{DL}\ell}{(2\pi)^{DL}} 
\frac{1}{S_i {\cal D}_i} \times \left\{
\begin{matrix}
(n_T-1)^{|i|}(N_i)^2\\ 
(n_V+D-6)^{|i|} N_i \tilde N_i
\end{matrix}\right. \,,
\end{equation}
which in $D=6$ yield the (2,0) and (1,1) theories, respectively. And for $D<6$, the numerators lose their chirality, $\tilde N_i = N_i$, and thus the double copy only give a single theory with $n_V$ vectors.  

The one-loop four-hyper amplitude is given in terms of the five diagram topologies in Fig.~\ref{fig:1loopBCJdiagrams}, and their 6D BCJ numerators can be worked out from \eqns{eqn11}{disco_grav}, or via 4D results \cite{Kalin:2018thp,Ben-Shahar:2018uie}, to be
\begin{equation} \label{OneLoopNumerators}
    N_{\rm (a)}=-N_{\rm (b)}=N_{\rm (c)}=-\frac{1}{2}N_{\rm (d)}=N_{\rm (e)} =s\,.
\end{equation}
As before, we have pulled out an overall supersymmetric $\delta^{(4)}(Q)$, which entirely removes the chiral nature of the numerators. Furthermore, the numerators are given for $USp(2)$ flavor, and for $USp(n_\Phi)$ one can dress them with skew-symmetric flavor factors following the arrows connecting external lines in  Fig.~\ref{fig:1loopBCJdiagrams}. In the double copy (\ref{DoubleCopyAnyL}), the two copies of symplectic groups are traded for $SO(n_T)$ or $SO(n_V)$, with flavor factors using $\delta^{f_if_j}$'s.

Since the one-loop numerators (\ref{OneLoopNumerators}) have no dependence on the loop momentum $\ell$, the 6D UV divergence comes entirely from the triangle and bubble integrals, and can be readily computed in the (2,0) supergravity theory as
\begin{align} \label{UVdiv}
 \redA^{\oneloop}_{TT}\Big|_{\rm UV}\!\!\!\!=& \Big[2 (N_{\rm (b)}^2\! {+} N_{\rm (c)}^2) I_3(s) {+}\frac{1}{2} \big(N_{\rm (d)}^2 {+} (n_T {-} 1) N_{\rm (e)}^2\big) I_2(s)\Big] \nonumber \\
 & \times \delta^{f_1f_2}\delta^{f_3f_4}+ {\rm cyclic}(1,2,3)\Big|_{\rm UV}\\
 =& \frac{s^2}{2}  (n_T-21) I_2(s) \delta^{f_1f_2}\delta^{f_3f_4}+ {\rm cyclic}(1,2,3)\,.\nonumber
\end{align}
Here $I_3(s)$ and $I_2(s)$ are the scalar triangle and bubble integrals (including $1/s$ propagators) in $D=6-2\epsilon$ dimensions, which are related by integration by parts as
\begin{equation}
I_3(s)=-\Big(3+\frac{\epsilon}{1-\epsilon}\Big)I_2(s)=\frac{1}{\epsilon}\frac{1}{2 (4\pi)^3 s}+{\cal O}(\epsilon^0)\,.
\end{equation}
The non-chiral (1,1) supergravity divergence, $ \redA^{\oneloop}_{VV}\big|_{\rm UV}$, is given by the same expression (\ref{UVdiv}), except that the factor $(n_T {-} 1)$ is replaced by $n_V$, as follows from setting $D=6$ in \eqn{DoubleCopyAnyL}. Thus, again, we see that the (2,0) and (1,1) UV divergences vanish for $n_T=21$ and $n_V=20$, respectively; see also comment~\footnote{For $D=4$, only the bubble integral is divergent, giving a UV-divergence proportional to $[4+(n_V+D-6)]/2 = (1+n_V/2)$, in agreement with \rcite{Fischler:1979yk} after accounting for $m=n_V-2$ in that paper.}.

We can also compute the local polynomial remainder of the dimensionally regulated amplitude, by expanding the first line of \eqn{UVdiv} to finite order ${\cal O}(\epsilon^0)$. For either $n_T=21$ or $n_V=20$ it gives the same answer:
\begin{equation}
\hat R=-\frac{2}{3(4\pi)^3}\Big( \delta^{f_1 f_2}\delta^{f_3 f_4}s + \delta^{f_1 f_4}\delta^{f_2 f_3} t + \delta^{f_1 f_3}\delta^{f_2 f_4} u \Big) \,. 
\end{equation}
The complete finite amplitudes are 
$\delta^{(8)}(Q)(\redA_{TT}^{\oneloop} + \hat R)$ and $\delta^{(4)}(Q)\delta^{(4)}(\tilde Q)(\redA_{VV}^{\oneloop} + \hat R)$ in dimensional regularization.

\begin{figure}[t]
  \centering
  \begin{subfigure}[a]{0.082\textwidth}
    \centering
\includegraphics[width=\textwidth]{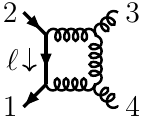}   
    \caption{}
  \end{subfigure}
  \hfill  
  \begin{subfigure}[a]{0.082\textwidth}
    \centering
\includegraphics[width=\textwidth]{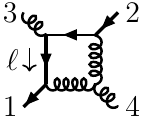}   
    \caption{}
  \end{subfigure}
  \hfill
  \begin{subfigure}[a]{0.082\textwidth}
    \centering
\includegraphics[width=\textwidth]{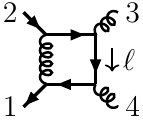}
    \caption{}
  \end{subfigure}
    \hfill
      \begin{subfigure}[a]{0.098\textwidth}
    \centering
\includegraphics[width=\textwidth]{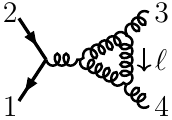} 
    \caption{}
  \end{subfigure}
\hfill  
\begin{subfigure}[a]{0.098\textwidth}
    \centering
\includegraphics[width=\textwidth]{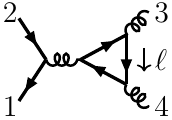} 
    \caption{}
  \end{subfigure}
  \caption{The BCJ diagram topologies for 6D SQCD with external mixed gauge and hypermultipets.}
  \label{fig:1loopBCJdiagramsMixed}
\end{figure}

\vspace{3pt}

{\bf Matter-graviton divergence.}
The one-loop BCJ diagrams for the SQCD amplitude with mixed hyper and gauge multiplets are given in Fig.~\ref{fig:1loopBCJdiagramsMixed}. The corresponding numerators (for $USp(2)$ flavor) can be worked out via unitarity, or by uplifting 4D results \cite{Kalin:2018thp}, to be
\begin{equation} \label{MixedOneLoopNumerators}
    N_{\rm (a)}\!=N_{\rm (b)}\!=N_{\rm (c)}\!=-\frac{1}{2}N_{\rm (d)}\!=N_{\rm (e)}\!=[3_{\hat a_3}|\ell|4_{\hat a_4}]\,, 
\end{equation}
where we have removed the $\delta^{(4)}(Q)$ factor. The numerators are explicitly chiral and linear in the loop momentum. The double copy thus yields rank-2 tensor integrals.  

The $D=6$ UV divergence of rank-2 tensor box integrals can be computed using the simple reduction formula $\ell^\mu\ell^\nu\to \ell^2 \eta^{\mu\nu}/D$.
In contrast, the rank-2 tensor triangles, which are formally power divergent, need a more sophisticated treatment to extract the logarithmic divergence, which results in the reduction rule
\begin{equation}
\ell^\mu\ell^\nu\to - \frac{2s\, \eta^{\mu \nu}}{D (D+2)}+\ldots\,.
\end{equation}
The ellipsis are terms built out of the vectors $p_3^\mu,p_4^\mu$, which will be killed by the on-shell spinors. 
Applied to the doubled numerators, the tensor reduction gives
\begin{align}\label{numer_dc}
(N_i)^2&  \to c_i [3_{(\hat a_3}|\gamma_\mu|4_{(\hat a_4}][3_{\hat b_3)}|\gamma^\mu|4_{\hat b_4)}] =0\,, \\
N_i\tilde N_i&  \to  c_i [3_{\hat a_3}\!|\gamma_\mu|4_{\hat a_4}]\langle 3_{b_3}\!|\gamma^\mu|4_{b_4}\rangle \! =\! {-}2 c_i \langle 3_{b_3}|4_{\hat a_4}]\langle 4_{b_4}|3_{\hat a_3}], \nonumber
\end{align}
where $\langle i_{a}| j_{\hat{b}}] = \lambda^A_{i, a} \tilde{\lambda}_{j, A \hat{b}}$ and the reduction coefficients are $c_{\rm (a)}=c_{\rm (b)}=c_{\rm (c)}=\ell^2/6$ and $c_{\rm (d)}= 4 c_{\rm (e)}=-s/6$.
The (2,0) theory has no mixed tensor-graviton UV divergence (for any value of $n_T$) since the chiral numerators integrate to spinor invariants that vanish identically.

For the (1,1) theory we get the following formula for the mixed vector-graviton UV divergence:
\begin{align} \label{UVdivMixed}
& \redA^{\oneloop}_{VH}\Big|_{\rm UV} \!=-2 \langle 3_{b_3}|4_{\hat a_4}]\langle 4_{b_4}|3_{\hat a_3}]\frac{s}{6}\Big[6-\frac{4+n_V}{4} \Big]   I_3(s) \delta^{f_1f_2}~~ \nonumber\\
 &~~~~~~~ = \langle 3_{b_3}|4_{\hat a_4}]\langle 4_{b_4}|3_{\hat a_3}]\frac{s}{12}   (n_V-20)  I_3(s) \delta^{f_1f_2}\,.
\end{align}
Again, we see a surprising cancellation for $n_V=20$. 

The above results, together with the finiteness of the four-graviton amplitude arising from the absence of allowed counterterms, show that the $\mathcal{N}=(2,0)$ and $\mathcal{N}=(1,1)$ theories are finite at one loop for $n_V=20$ and $n_T=21$, respectively. This is 
rather unexpected, since symmetry-allowed counterterms generically do exist. For the four-matter amplitude in the $(2,0)$ theory, the relevant counterterm is proportional to
\begin{equation}\label{eq: CT4pt}
\delta^{(8)}(Q)\Big( \delta^{f_1 f_2}\delta^{f_3 f_4}s + \delta^{f_1 f_4}\delta^{f_2 f_3} t + \delta^{f_1 f_3}\delta^{f_2 f_4} u \Big) \, ,
\end{equation}
and an analogous expression holds for the $(1,1)$ theory, with $\delta^{(8)}(Q)$ replaced by $\delta^{(4)}(Q)\delta^{(4)}(\tilde Q)$.
For the mixed graviton–matter amplitude, \eqn{numer_dc} shows that in the $\mathcal{N}=(2,0)$ theory the only symmetry-preserving counterterm vanishes; while in the $\mathcal{N}=(1,1)$ theory, it is given by
\begin{equation}
\delta^{(4)}(Q)\delta^{(4)}(\tilde{Q})\delta^{f_1f_2}\langle 3_{a_3}|4_{\hat{a}_4}]\langle 4_{b_4}|3_{\hat{a}_3}]\, .
\end{equation}
Finally, note that the presence of divergences for generic values of $n_V$ and $n_T$, suggests that the counterterms cannot be ruled out based on duality symmetries of the scalars in the multiplet, analogous to the analysis in \rcites{Elvang:2010kc, Beisert:2010jx, Kallosh:2023dpr}.

\section{Conclusions}
In this Letter we studied the UV behavior of four-point amplitudes at one-loop order in 6D $(2,0)$ supergravity coupled to $n_T$ tensor multiplets, as well as in  6D $(1,1)$ supergravity coupled to $n_V$ vector multiplets, using unitarity and double-copy-based approaches.
Interestingly, we found that all these one-loop amplitudes are UV-finite when the numbers of multiplets are tuned to $n_T=21$ and $n_V=20$, respectively. Remarkably, these values coincide precisely with those emerging from type II string theories compactified on K3.

Several open questions remain, which we hope to address in the near future. Firstly, it is crucial to understand the underlying reason for the UV finiteness. Since generic supergravity theories do not exhibit such cancellations, something special seems to be at play in 6D.
 As noted in the introduction, it is tempting to speculate that $T$-duality plays a role in forcing the two amplitudes to coincide when $n_T=21$ and $n_V=20$. It would be interesting to explore this idea further and establish whether such a connection can be made precise.

Secondly, it was shown in  \rcites{Beccaria:2015uta, Casarin:2024qdn} that 6D $(2, 0)$ conformal supergravity coupled to $26$ tensor multiplets is free of the $a$-anomaly. It was suggested~\cite{Beccaria:2015uta} that by treating five of these tensor multiplets as ghost-like compensators and spontaneously breaking conformal symmetry, one obtains the 6D $(2, 0)$ supergravity with $21$ tensor multiplets as the low-energy theory. It would be of great interest to understand whether there exist further direct connections between these two (2,0) theories, and what are the implications at higher loop orders. 

Furthermore, it is essential to investigate whether the UV finiteness extends beyond one loop and to higher-point amplitudes involving external gravitons or tensors, building on the tree-level amplitudes of \rcite{Heydeman:2018dje}. For the non-chiral (1,1) theory, the result in \rcite{Bern:2013qca} indicates that the four-point two-loop amplitude does diverge. An analogous analysis of UV divergences in the (2,0) theory has not been performed yet, we hope to report on this in the near future.

It would also be worthwhile to consider supergravity theories with matters in other spacetime dimensions. 
Subdivergences in even dimensions make two-loop calculations more subtle and beyond the scope of this work. However, we have verified, using the same double-copy approach as in \rcite{Bern:2012gh}, that in $D=5$ there exists a similar cancellation of the two-loop four-point UV divergence in half-maximal supergravity with exactly $n_V = 5$ vector multiplets, for arbitrary external matter and graviton states,  in agreement with \cite{Bern:2013qca}. 
Note that this theory can be obtained from 10D $\mathcal{N}=1$ supergravity reduced on $T^5$.
It is worth noting that a recent study \cite{Chester:2023qwo} also observed an interesting cancellation of two-loop logarithmic divergences in 7D arising from M-theory on orbifolds \footnote{This cancellation results from an interplay between two-loop gluon contributions confined to 7D and tree-level graviton exchange in 11D.}.

Finally, it would be extremely interesting to see if the special features observed here survive when the theory is considered in AdS background. For weakly coupled 6D (2,0) supergravity on AdS$_3 \times$S$^3$ with Ramond-Ramond fluxes, the dual CFT is the D1–D5 system in its strong-coupling regime \cite{Maldacena:1997re}. Since all four-point tree-level correlators have been obtained \cite{Giusto:2019pxc, Rastelli:2019gtj,Giusto:2020neo,Aprile:2021mvq,Wen:2021lio,Behan:2024srh,Aprile:2025kfk}, one can apply analytic bootstrap techniques to compute one-loop correlators and investigate aspects of $n_T=21$ in an AdS setting \cite{Aprile:2026yit}.

\section{Acknowledgments}
We are grateful to Lance Dixon, Oliver Schlotterer, John Schwarz, and Chia-Hsien Shen for enlightening discussions. MS would like to thank Francesco Aprile and Hynek Paul for many discussions and collaboration on related topic. 
Y-tH is supported by the Taiwan National Science and Technology Council grant 112-2628-M-002-003-MY3 and 114-2923-M-002-011-MY5.
HJ is supported by the Knut and Alice Wallenberg Foundation under
grants KAW 2018.0116 and KAW 2018.0162.
MS is funded by the European Union (ERC Synergy Grant MaScAmp 101167287). Views and opinions expressed are however those of the author(s) only and do not necessarily reflect those of the European Union or the European Research Council Executive Agency. Neither the European Union nor the granting authority can be held responsible for them. 
CW is supported by a Royal Society University Research Fellowship,  URF$\backslash$R$\backslash$221015 and a STFC Consolidated Grant, ST$\backslash$T000686$\backslash$1 ``Amplitudes, strings \& duality".

\onecolumngrid

\resumetoc
\newpage
\vspace{2cm}

\setcounter{equation}{0} % Reset equation counter
\renewcommand{\theequation}{S\arabic{equation}} % Redefine equation numbering
\setcounter{secnumdepth}{2}

\begin{center}
\textbf{\large Supplemental Material}
\end{center}

%\tableofcontents

\parskip0.3cm

%================================================================================================
\section{Phase-space and Feynman integrals}
%================================================================================================
In this section, we collect the results of several phase-space integrals, and a few Feynman integrals, that arise in the computation of one-loop amplitudes.

The phase-space integral is normalized so that
\begin{equation}\label{normal_phasespace}
\int  d\omegatwo = -  \frac{1}{2(4\pi)^3} \int_0^\pi  d\phi_{\text{fn}}  \int_0^\pi d\cos\theta_{\text{in}} \sin^2{\theta_{\text{in}}} \sin^2{\phi_{\text{fin}}} =   \frac{1}{192\pi^2} \,.
\end{equation}
A convenient basis of polynomials, orthogonal with respect to the phase-space measure, is provided by the partial waves. In $D$ dimensions, they take the form
\begin{equation}
P_{\ell}^{(\frac{D-3}{2})}(z) =2^{2D-5} \pi^{\frac{D-3}{2}}  (2\ell+D-3)\Gamma \Big(\frac{D-3}{2}\Big) C_\ell^{(\frac{D-3}{2})}(z)\,,
\end{equation}
with $C_\ell^{(\frac{D-3}{2})}(z)$ the Gegenbauer polynomials.
In the normalization conventions of \eqref{normal_phasespace}, the 6D partial waves satisfy the orthonormality condition
\begin{equation}\label{orto_pw}
\frac{1}{s} \int  d\omegatwo P_{\ell}^{(\frac{3}{2})} (\cos \theta_{\text{fn}})P_{\ell'}^{(\frac{3}{2})}(\cos \theta_{\text{in}}) = \delta_{\ell, \ell'} P_{\ell}^{(\frac{3}{2})}(\cos \theta) \,.
\end{equation}
The partial-wave expansion offers an alternative method for computing discontinuities of scalar scattering amplitudes. In short, one expands the tree-level amplitudes on either side of the cut in partial waves, and, due to the orthonormality property \eqref{orto_pw}, the phase-space integral reduces to a (possibly infinite) sum over partial waves. Further details can be found in \rcite{Huang:2024ihm}.

{\bf{Tensor cut.}} The relevant integrals for the tensor cut are
\begin{align}
    & \int  \frac{1}{s_{1\ell_2}s_{3\ell_1}} d\omegatwo= \int  \frac{1}{s_{1\ell_1}s_{3\ell_2}} d\omegatwo =  -\frac{1}{32\pi^2}  \frac{1}{s u} \log \Big(-\frac{s}{t}\Big) \,,   \notag \\[.5cm]
& \int  \frac{1}{s_{1\ell_1}s_{3\ell_1}} d\omegatwo = \int  \frac{1}{s_{1\ell_2}s_{3\ell_2}} d\omegatwo =  -\frac{1}{32\pi^2}  \frac{1}{s t} \log \Big(-\frac{s}{u}\Big) \,,   \notag  \\[.5cm]
& \int \Big(  \frac{1}{s_{1\ell_1}}+ \frac{1}{s_{1\ell_2}}+ \frac{1}{s_{1\ell_1}}+ \frac{1}{s_{3\ell_2}} \Big) d\omegatwo =  \frac{1}{16\pi^2 s} \, .
\end{align}

{\bf{Graviton cut.}} For the graviton cut in the (2,0) theory, we need to evaluate \eqn{disco_grav}. Using the trace identity
\begin{equation}
    \Tr (\gamma^\mu \gamma^\nu \gamma^\rho\gamma^\sigma) = 4 (\eta^{\mu\nu}\eta^{\rho\sigma}+\eta^{\mu\sigma}\eta^{\nu\rho} - \eta^{\mu\rho}\eta^{\nu\sigma}) \,,
\end{equation}
we arrive at
\begin{equation}
 \int \Big( \frac{s^2 u^2}{s_{1\ell_1}s_{3\ell_1}s_{1\ell_2}s_{3\ell_2}} {+} \frac{2 s u}{s_{1\ell_2}s_{3\ell_1}} {+}\frac{2 s u}{s_{1\ell_1}s_{3\ell_2}} {+}\frac{s_{1\ell_2}s_{3\ell_1}}{s_{1\ell_1}s_{3\ell_2}}{+} \frac{s_{1\ell_1}s_{3\ell_2}}{s_{1\ell_2}s_{3\ell_1}}+1 \Big) d\omegatwo \,.
\end{equation}
On the other hand, for the (1,1) theory, we get
\begin{equation}
 \int \frac{(s_{1\ell_2} s_{3\ell_1}+ s_{1\ell_1} s_{3\ell_2} + s u)^2}{s_{1\ell_1}s_{3\ell_1}s_{1\ell_2}s_{3\ell_2}} d\omegatwo \,.
\end{equation}

Compared to the four-tensor integrals, the only new integrals appearing are
\begin{align}
    & s^2 u^2 \int \frac{1}{s_{1\ell_1}s_{3\ell_1}s_{1\ell_2}s_{3\ell_2}}d\omegatwo =    -\frac{1}{16\pi^2 }\Big( \frac{u^2}{s t} \log \Big(-\frac{s}{u}\Big) + \frac{u}{s} \log \Big(-\frac{s}{t}\Big) \Big) \,,  \notag \\[.5cm]
 &  \int \frac{s_{1\ell_2}s_{3\ell_1}}{s_{1\ell_1}s_{3\ell_2}} d\omegatwo = \int \frac{s_{1\ell_1}s_{3\ell_2}}{s_{1\ell_2}s_{3\ell_1}} d\omegatwo \, = -\frac{1}{32\pi^2 } \Big( \frac{s}{u} \log \Big(-\frac{s}{t}\Big) +\frac{5}{6} \Big)\,.
\end{align}
Summing all contributions, we arrive at the result quoted in the main text, namely
\begin{equation}
 \int \Big( \frac{s^2 u^2}{s_{1\ell_1}s_{3\ell_1}s_{1\ell_2}s_{3\ell_2}} {+} \frac{2 s u}{s_{1\ell_2}s_{3\ell_1}} {+}\frac{2 s u}{s_{1\ell_1}s_{3\ell_2}} {+}\frac{s_{1\ell_2}s_{3\ell_1}}{s_{1\ell_1}s_{3\ell_2}}{+} \frac{s_{1\ell_1}s_{3\ell_2}}{s_{1\ell_2}s_{3\ell_1}}+1 \Big) d\omegatwo =  -\frac{1}{16\pi^2 } \Big( \frac{t^2}{s u} \log \Big(-\frac{s}{t} \Big)+\frac{u^2}{s t} \log \Big(-\frac{s}{u} \Big) + \frac{3}{4}  \Big) \,
\end{equation}
for the (2,0) theory and
 \begin{equation}
 \int \frac{(s_{1\ell_2} s_{3\ell_1}+ s_{1\ell_1} s_{3\ell_2} + s u)^2}{s_{1\ell_1}s_{3\ell_1}s_{1\ell_2}s_{3\ell_2}} d\omegatwo \, =  -\frac{1}{16\pi^2 } \Big( \frac{t^2}{s u} \log \Big(-\frac{s}{t} \Big)+\frac{u^2}{s t} \log \Big(-\frac{s}{u} \Big) + \frac{2}{3}  \Big) \,
\end{equation}
for the (1,1) theory.

{\bf{Dimensional regularization.}}  The dimensionally regulated one-loop bubble and triangle scalar integrals are defined as
 \begin{align}
 I_2(s)&=\frac{i}{s^2}\int\frac{d^D\ell}{(2\pi)^D}\frac{1}{\ell^2 (\ell+p_1+p_2)^2}=-\frac{(-s)^{D/2-4}}{(4\pi)^{D/2}}\frac{\Gamma(2-D/2)\Gamma^2(D/2-1)}{\Gamma(D-2)}\,, \nonumber \\
 I_3(s)&=\frac{i}{s}\int\frac{d^D\ell}{(2\pi)^D}\frac{1}{\ell^2 (\ell+p_1)^2 (\ell+p_1+p_2)^2}= -2 \frac{D - 3}{D - 4}I_2(s)\,,
 \end{align}
 where $s=(p_1+p_2)^2$ and $p_i^2=0$. For $D=6$, we give the massless ($p_i^2=0$) scalar box integral as
\begin{align}
  B(s,t)&=\frac{1}{(4 \pi)^3} {\cal B}_{s,t} =i\int\frac{d^6\ell}{(2\pi)^D}\frac{1}{\ell^2 (\ell+p_1)^2 (\ell+p_1+p_2)^2 (\ell+p_1+p_2+p_3)^2}=\frac{1}{2(4 \pi)^3}\frac{1}{s+t}\Big(\log^2\frac{s}{t}+\pi^2\Big)\,, 
\end{align}
where the analytic answer is valid in the unphysical region $s/t>0$. 

%%%%%%%%%%%%%%%%%%%

\twocolumngrid

\bibliographystyle{apsrev4-2}

\end{document}